\documentclass[useAMS,usenatbib]{mn2e}
\usepackage{graphicx}
\usepackage{amsmath}
\usepackage{makecell}
\usepackage{subcaption}
\usepackage{gensymb}
\usepackage[normalem]{ulem}
\usepackage{macros}
\usepackage[usenames,dvipsnames,svgnames,table]{xcolor}
\usepackage{hyperref}
\definecolor{darkblue}{rgb}{0.0,0.0,0.3}
\hypersetup{colorlinks,%
            linkcolor=darkblue,urlcolor=darkblue,
            anchorcolor=darkblue,citecolor=darkblue}
\usepackage{txfonts}
\DeclareSymbolFont{cmletters}{OML}{cmm}{m}{it}
\DeclareMathSymbol{v}{\mathalpha}{cmletters}{"76}
\newcommand{\RedeclareMathOperator}[2]{\renewcommand{#1}{}\let#1\relax\DeclareMathOperator{#1}{#2}}

\newcommand\simless\lesssim
\newcommand\simgreat\gtrsim

\def\hide#1{}

\title[A Phase Lag Between Disc and Corona in GRMHD Simulations of Precessing Tilted Accretion Discs]{A Phase Lag Between Disc and Corona in GRMHD Simulations of Precessing Tilted Accretion Discs}
\author[Liska, Hesp, Tchekhovskoy, Ingram, van der Klis \& Markoff]{M. Liska$^{1}$, C. Hesp$^4$, A. Tchekhovskoy$^{2}$, A. Ingram$^3$, M. van der Klis$^4$ \& S.B. Markoff$^4$ \\
$^{1}$Institute for Theory and Computation, Harvard University, 60 Garden Street, Cambridge, MA 02138, USA \\
$^{2}$Center for Interdisciplinary Exploration \& Research in Astrophysics (CIERA), Physics \& Astronomy, Northwestern University, Evanston, IL 60202, USA\\
$^{3}$School of Mathematics, Statistics and Physics, Newcastle University, Herschel Building, Newcastle Upon Tyne, UK \\
$^{4}$Anton Pannekoek Institute for Astronomy, University of Amsterdam, Science Park 904, 1098 XH Amsterdam, The Netherlands
}
\begin{document}

\date{Accepted. Received; in original form}
\pagerange{\pageref{firstpage}--\pageref{lastpage}} \pubyear{2017}

\maketitle
\label{firstpage}

\begin{abstract}
In the course of its evolution, a black hole (BH) accretes gas from a wide range of directions. Given a random accretion event, the typical angular momentum of an accretion disc would be tilted by $\sim$60$^\circ$ relative to the BH spin. Misalignment causes the disc to precess at a rate that increases with BH spin and depends on disc morphology. We present general-relativistic magnetohydrodynamic (GRMHD) simulations spanning a full precession period of highly tilted (60$^\circ$), moderately thin ($h/r=0.1$) accretion discs around a rapidly spinning ($a\simeq0.9$) BH. While the disc and jets precess in phase, we find that the disc wind/corona, sandwiched between the two, lags behind by $\gtrsim 10^{\circ}$. For spectral models of BH accretion, the implication is that hard non-thermal (corona) emission lags behind the softer (disc) emission, thus potentially explaining some properties of the hard energy lags seen in Type-C low frequency quasi-periodic oscillations in X-Ray binaries. While strong jets are unaffected by this disc-corona lag, weak jets can stall when encountering the lagging corona at distances $r \sim 100$ BH radii. This interaction may quench large-scale jet formation. 

\end{abstract}


\section{Introduction} \label{sec:introduction} 

Evidence is growing that many accreting black holes (BHs) have accretion discs that are misaligned relative to the BH equator. Because supermassive BH spin magnitude and direction are set by the history of randomly oriented gas accretion events and galaxy mergers, misaligned accretion can naturally arise \citep{Volonteri2005,King2005}. Indeed, misalignment between inner and outer disc is found in AGN masers (see e.g. \citealt{Herrnstein2005,Caproni2006,Caproni2007, Greene2013}). For X-ray binaries (XRBs) the dynamical evolution of the system, which may include asymmetric supernova kicks and 3-body interactions in stellar clusters, can also lead to substantial misalignment. Indeed, the jets in XRB GRO~J1655--40 are offset by $15^\circ$ from the binary plane (\citep{hjellming95,Greene2001,Maccarone2002}).

In view of these observations, it is important to understand how accretion discs respond to the misalignment relative to the BH spin. An important effect here is general relativistic (GR) frame dragging, which is associated with the BH spin and known to induce nodal \citet[LT hereafter]{lense18} precession of orbits inclined to the BH equatorial plane. As the precession frequency decreases with distance from the BH roughly as $\Omega_{LT}\propto a/r^{3}$, frame dragging is predicted to warp tilted accretion discs. When the disc is geometrically thick, i.e. the dimensionless scale height is larger than the viscosity parameter, $h/r>\alpha$, as would be expected in all low luminosity AGN and hard state XRBs \citep{Narayan1994}, viscous diffusion of the warp plays a minimal role \citep{Papaloizou1983} and the warp is communicated radially through bending waves traveling at approximately half the speed of sound \citep{papaloizou95}. When the tilt (measured in radians) is small compared to $h/r$, a bending wave dominated disc accretes misaligned and forms smooth radial oscillations in tilt closer to the BH \citep{Ivanov1997, Lubow2002, Fragile2007, Liska2018A}. Since a larger non-linear tilt, exceeding the disc scale height, is expected in many systems \citep{King2005}, it is crucial to understand how a disc responds to such a large tilt.

General relativistic magnetohydrodynamic (GRMHD) simulations are an excellent tool to gain insight into such problems because they self-consistently model the anisotropic (turbulent) stresses \citep{Balbus1991,Balbus1998} responsible for (mis)aligned angular momentum transport and warp propagation. Solid-body precession of tilted accretion discs has been suggested \citep{Stella1998, Fragile2007} as the origin of low frequency quasi-periodic oscillations (QPOs) observed in XRB lightcurves (e.g., \citep{Klis1989}). However, it is inconsistent with the observed energy dependence of QPO frequency (e.g. \citealt{Eijnden2016A,Eijnden2017}), suggesting that a more complex geometry, involving differential precession between various disc and/or jet components, may be present.

 Here we present the first GRMHD simulations of highly tilted accretion discs in the bending wave, $h/r \simgreat \alpha$, regime, that show signs of a phase lag between different components of the accretion flow. We describe our numerical setup in Sec.~\ref{sec:numerical-models}, present our results in Sec.~\ref{sec:Results}, and conclude in Sec.~\ref{sec:Conclusion}.

\section{Numerical models}
\label{sec:numerical-models}
We use a massively parallel 3D GRMHD code H-AMR \citep{Liska2018A, Liska2020} accelerated by Graphical Processing Units (GPUs). We use a spherical-polar grid of resolution $1372\times480\times738$, which is uniform in $\log r$ and $\phi$ and approximately uniform in $\theta$ (the grid is slightly stretched out in $\theta$ within $\sim 5^{\circ}$ of the pole as described in \citep{Ressler2015}). This sufficiently resolves the fastest growing wavelength of the magnetorotational instability (MRI; \citep{Balbus1991}). Here the quality factors $Q^{i}=\lambda_{mri}^{i}/N^i$ (as defined in \citep{Liska2018C}), which give the number of cells ($N^i$) per MRI-wavelength ($\lambda_{mri}^{i}$) in the $i-$th dimension, exceed ($Q^{r} \times Q^{\theta} \times Q^{\phi}) \ge (10 \times 10 \times 40$) at $t=4 \times 10^4 r_g/c$. This guarantees a reasonable degree of convergence (e.g., \citep{2013ApJ...772..102H}). To prevent the cells from becoming elongated near the poles, we reduce the $\phi$-resolution sequentially by factors of two within $30^{\circ}, 15^{\circ}$, and $7.5^{\circ}$ away from the pole. We use outflow boundary conditions in the $r$-direction, with the inner radial boundary inside the BH event horizon and the outer boundary at $10^5r_g$, where $r_g=GM/c^2$ is the gravitational radius, such that both of the boundaries are causally disconnected from the flow. We apply transmissive boundary conditions in the $\theta$-direction ensuring free passage of the plasma through the polar singularity \citep{Liska2018A}. 

We initialize the simulations with an equilibrium torus \citep{Fishbone1976} around a BH with spin $a=0.9375$ in a Kerr-Schild foliation. We place the torus inner edge at $r_{\rm in} = 12.5r_g$, its density maximum at $r_{\rm max} = 25r_g$ (without loss of genereality we set $\rho_{\rm max}=1$). We use an ideal gas equation of state, $p_{g}=(\Gamma-1)u_{g}$, where $p_{g}$ and $u_{g}$ are gas thermal pressure and energy densities, and non-relativistic $\Gamma = 5/3$.  We tilt the initial torus relative to the BH spin axis (and grid) by an angle $\mathcal{T}_{\rm init} = 60\degree$ (see \citep{Liska2018A} for details). We cool the disc to its target thickness, $h/r=0.1$, on the Keplerian timescale using a source term \citep{Noble2009,Liska2018C}.
\begin{figure}
  \begin{center}
    \includegraphics[width=\linewidth,trim=0mm
    0mm 0mm 0,clip]{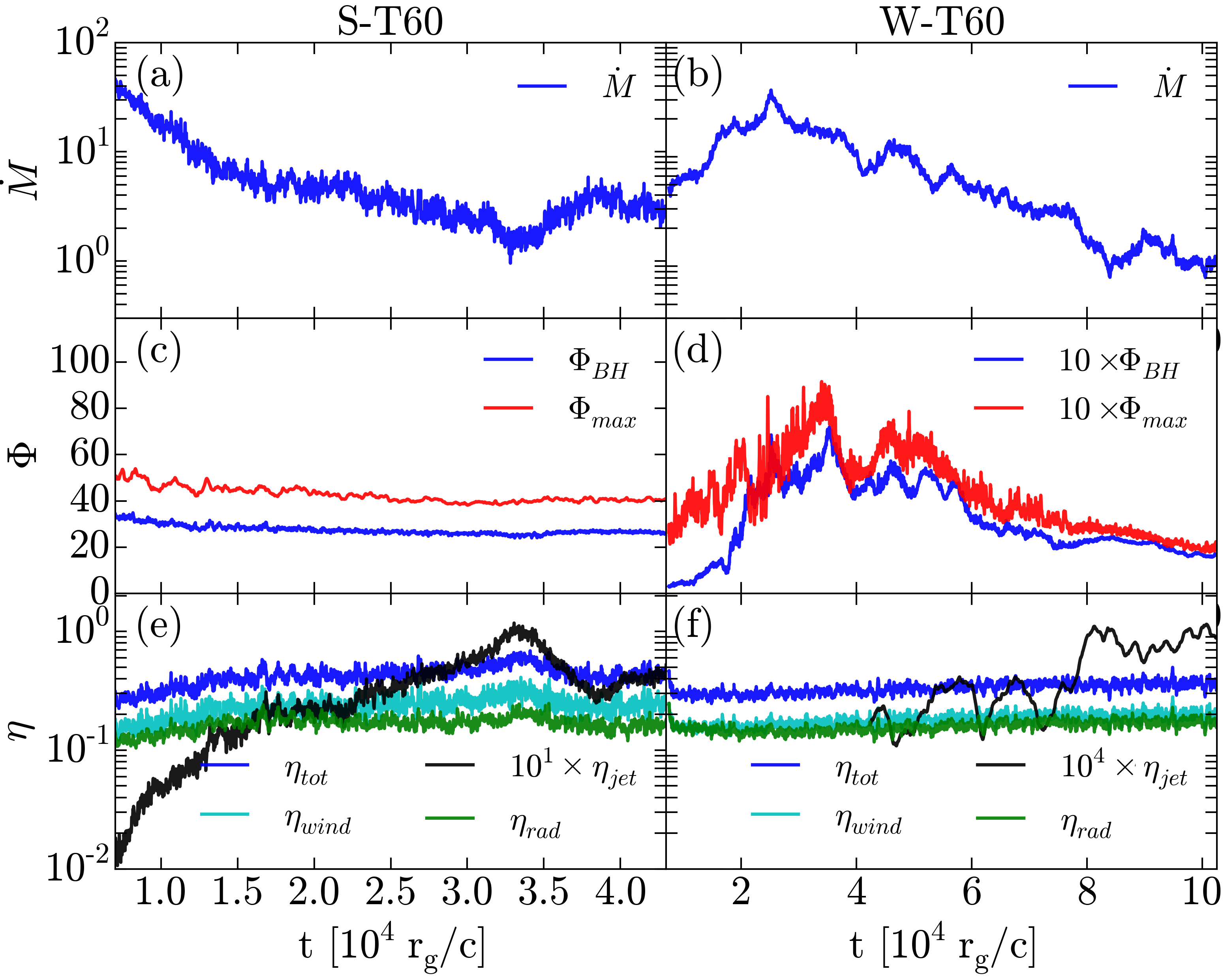} \end{center}
  \caption{The mass accretion rate ($\dot{M}$) in (a,b) and magnetic fluxes in the disc ($\Phi_{max}$) and on the BH ($\Phi_{BH}$) in (c,d) as function of time. While in (e,f) the radiative ($\eta_{rad}$) and wind ($\eta_{wind}$) efficiency are similar for both models, the jet efficiency ($\eta_{jet}$) is multiple orders of magnitude higher in S-T60. This can be explained by the presence of large scale magnetic flux in S-T60.}
\label{fig:time}
\end{figure}

We consider two models, W-T60 and S-T60 respectively, whose discs are both tilted by $60^\circ$. W-T60 features a small poloidal field loop concentrated near the BH with vector potential $A_{\phi}\propto(\rho-0.05)$, and S-T60 has a much larger loop, $A_{\phi}\propto(\rho-0.05)^{2}r^{3}$. We initially normalize the field strength by setting $\max p_{\rm g}/\max p_{\rm b}=5$ for W-T60 and $\max p_{\rm g}/\max p_{\rm b}=100$ for S-T60 which keeps both models fully gas pressure dominated in the initial conditions. $\max p_{\rm g}$ and $\max  p_{\rm b}$ are the maximum gas and magnetic pressure in the disc. The size of the loop turns out to be more important than the normalization in setting the total amount of embedded poloidal magnetic flux \footnote{This is partly because the size enters the magnetic flux in the second power, as opposed to the magnetic field strength, which enters in the first power. This is also partly because for a larger loop the maximum of the magnetic pressure, which enters the normalization, occurs at a larger radius.} and with that the strength of the jet (see Sec.~\ref{sec:Results}).

\begin{figure*} \begin{center}
\includegraphics[width=0.9\linewidth,trim=0mm 0mm 0mm
0,clip]{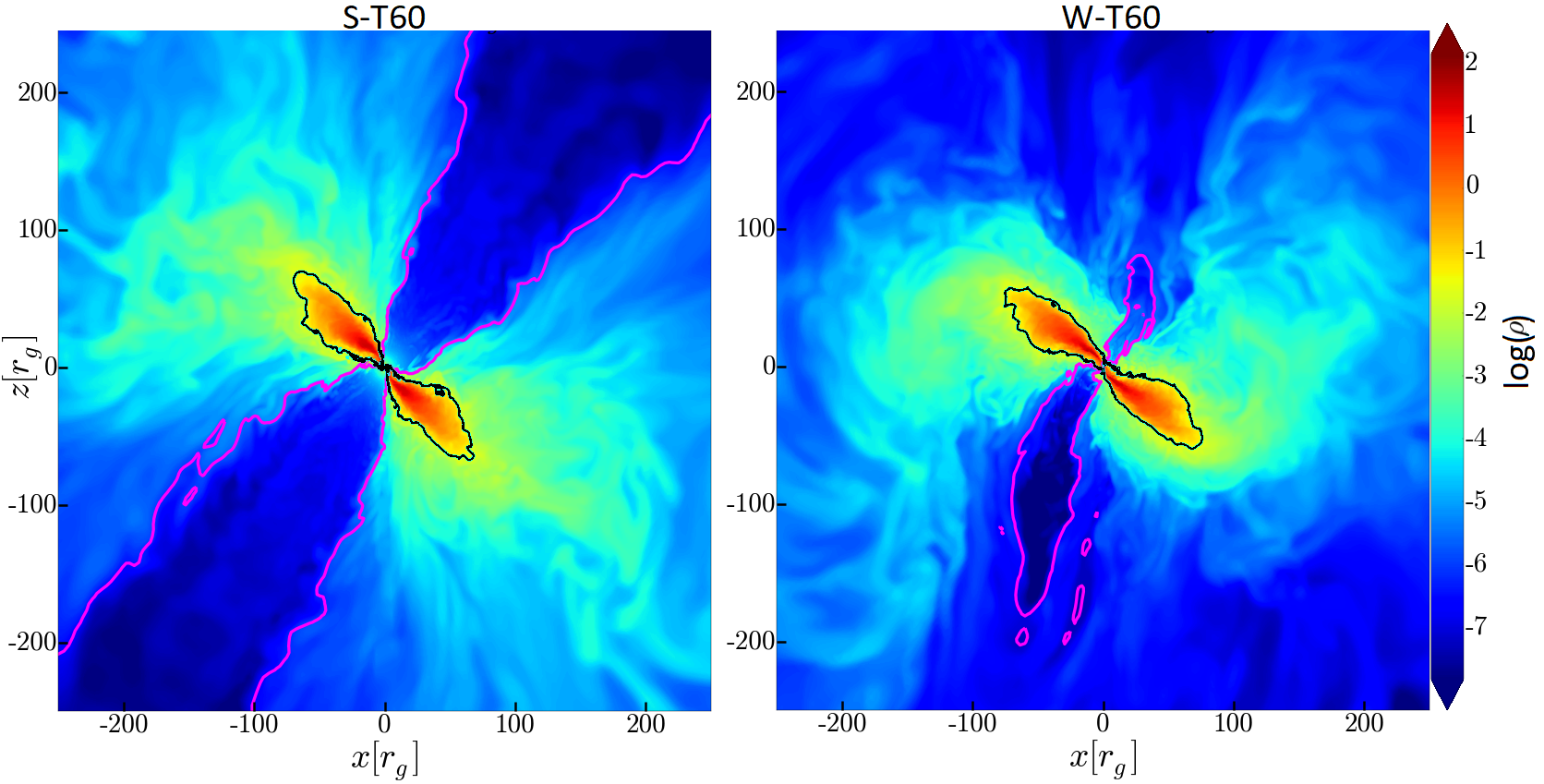} \end{center} \caption{A
colour map of the fluid-frame rest mass density $\log(\rho)$ for models S-T60 (left panel) and W-T60 (right panel)
at $t=4.5\times10^4$ $r_g/c$. The black hole spin axis is oriented along the vertical z-direction. Magenta lines show the corona-jet
boundary, and black lines the disc-corona boundary. The strong jet in model S-T60 (left panel ) readily accelerates to relativistic speeds (Fig.~\ref{fig:jetplot}). However, since the corona 
in model W-T60 precesses slower than the disc, it becomes misaligned relative to the disc and obstructs jet propagation. This causes the jet to stall beyond $\sim 10^2 r_g$ (Fig.~\ref{fig:jetplot}).}
\label{fig:contourplot} 
\end{figure*}

\section{Results}
\label{sec:Results}
Differential rotation shears up magnetic field lines, activating the MRI and leading to turbulence and accretion onto the BH.  We study the system at late time, $t\gtrsim 10^4 r_g/c$, after the disc has cooled down to its target thickness. Figure~\ref{fig:time}(a,b) shows that the mass accretion rate, $\dot{M}$, quasi-steadily declines in both models. Figure~\ref{fig:time}(c,d) shows the positive (there is no negative) poloidal magnetic flux on the BH,  $\Phi_{BH} = 0.5\int_{r=r_{\rm H}}\left|B^r\right| dA_{\theta\phi}$ and in the disc, $\Phi_{\rm max}=\max_{r} \Phi_{p}(r)$ with $\Phi_{p}(r)=\max_{\theta}\int_0^\theta B^r dA_{\theta\phi}$. Here $\max_{r}$ and $\max_{\theta}$ give the maximum value of the respective quantities for all $r$ and $\theta$. Compared to W-T60, model S-T60 contains an order of magnitude more flux, most of which threads the event horizon and may launch outflows.

Figure~\ref{fig:time}(e,f) breaks down the energy outflow efficiency $\eta_{tot}$, or the energy outflow rate divided by the BH rest-mass energy accretion rate, $\dot Mc^2$, into radiative, $\eta_{rad}$, mechanical outflow, $\eta_{wind}$, and jet, $\eta_{jet}$, constituents that we define below (see equations 6,7 and 9 in \citep{Liska2018C} for detailed definitions). The radiative efficiency, $\eta_{rad}$, approximated by integrating the total radiative cooling rate from the inner photon orbit at $r=1.43 r_g$ to $100 r_g$, hovers around the canonical \citet{Novikov1973} value of $18\%$. The jet efficiency, $\eta_{jet}$, corresponding to the mechanical energy carried by the jets, reaches peak values of around $0.01\%$ in W-T60 and $10\%$ in S-T60. The wind efficiency, $\eta_{wind}$, or the mechanical outflow efficiency outside of the jet, reaches around $20\%$ for both models at $r=5 r_g$. The sum is $\eta_{tot}\approx 40\%$. However, in both models $\eta_{wind}$ drops to zero at $r \sim 400 r_g$, suggesting that the wind launched at smaller radii is gravitationally bound and eventually circularizes within $r \lesssim 400 r_g$.

\begin{figure} \begin{center}
    \includegraphics[width=0.8\linewidth,trim=0mm 0mm 0mm
    0,clip]{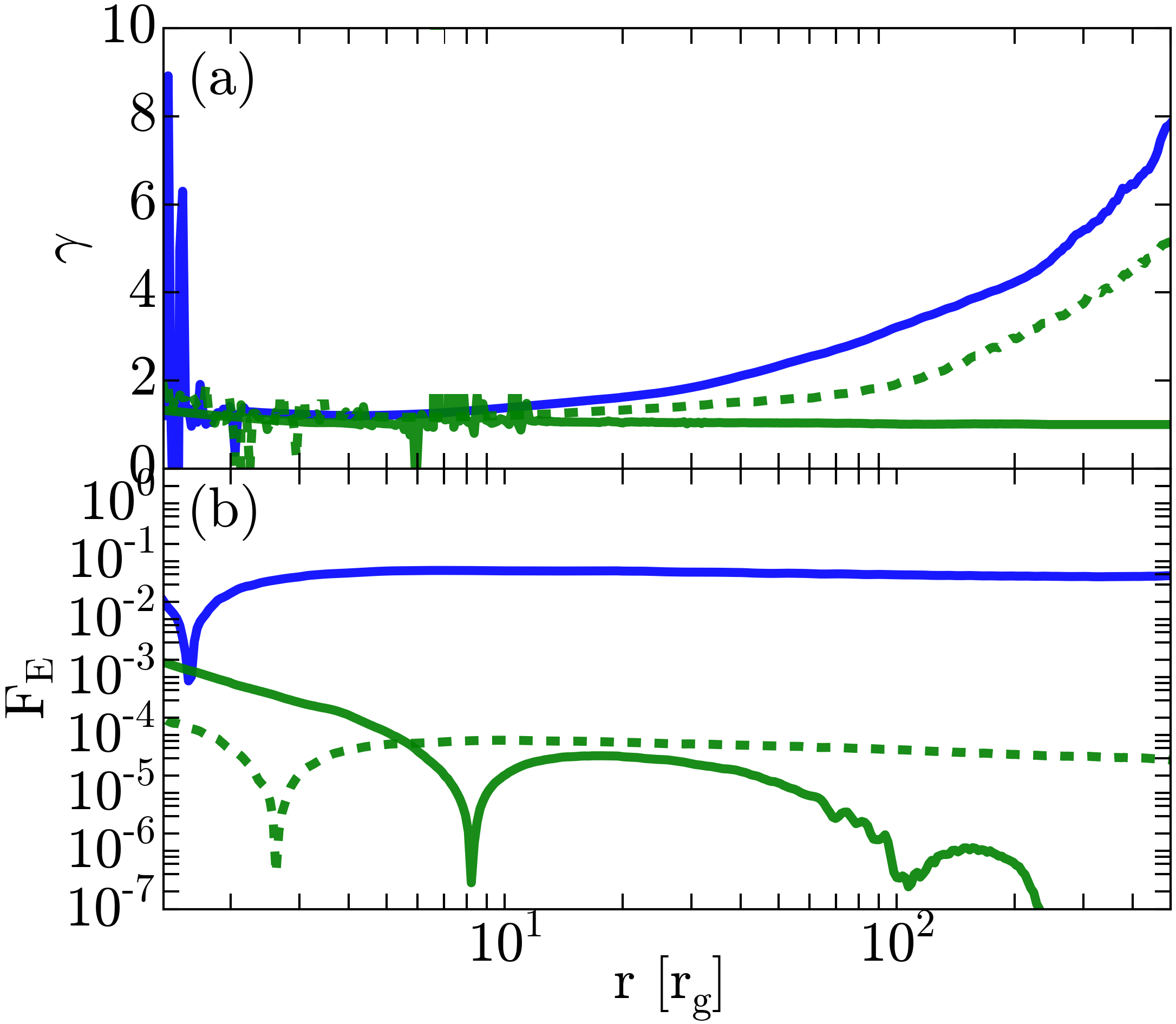} \end{center} \caption{Radial profiles of the Lorentz factor, $\gamma$, in panel (a), and energy flux, $F_E$, through the jet in panel (b). We plot those when the phase lag is maximum, at $t=2.0\times10^4 r_g/c$ for S-T60 (solid blue line) and at $t=4.2\times10^4 r_g/c$ (solid green) for W-T60. In addition, we show the radial profiles for W-T60 at $t=10\times10^4 r_g/c$ (dashed green) when the phase lag is small. The low Lorentz factor ($\gamma\sim1$) and drop of energy flux at $r\gtrsim50r_g$ in model W-T60 at early times indicate that a weak jet stalls and disrupts as it runs into the lagging corona. In contrast, acceleration to a relativistic Lorentz factor ($\gamma\sim7$) and near-constancy of the energy flux in model S-T60 indicate that a stronger jet survives this interaction. The constancy of the energy flux in model W-T60 at late times shows that for small tilt and phase lag values (see Fig.~\ref{fig:tiltvst}), even a weak jet can penetrate the precessing corona.}
\label{fig:jetplot}
\end{figure}

Figure~\ref{fig:contourplot} shows a colour map of the density $\rho$ in a vertical slice for models W-T60 and S-T60. We define the disc-corona boundary at $\rho=0.025\times \rho_{max}$ and the corona-jet boundary at $p_b/\rho c^2=5$. The corona-disc boundary was chosen to approximately match the $\beta=p_{g}/p_{b} \sim 1$ surface. This has the advantage over a more direct $\beta=1$ definition for the corona since it does not leave out coronal current sheets where $\beta >> 1$. Similar to recent GRMHD simulations \citep{Liska2022} the gas in the inner corona has a strong outflowing velocity component (see also \citep{Beloborodov1999, Fereira2006}). Note though that, as stated in the previous paragraph, this outflowing gas is gravitationnaly bound and eventually circularizes within $r \lesssim 400 r_g$. While the disc shapes appear to be similar in both models, the corona (yellow-green-cyan regions) in model W-T60 gets in the way of the jets and deflects them sideways, as seen in Fig.~\ref{fig:contourplot}(a) (see also SI or \href{https://www.youtube.com/playlist?list=PLDO1oeU33GwmwOV_Hp9s7572JdU8JPSSK}{our YouTube playlist}).  On the other hand, the stronger jet in model S-T60 propagates unhindered through the corona, pushing it aside. Below, we show that this elongation arises through a phase lag between the disc and corona. Figure~\ref{fig:jetplot} shows the Lorentz factor, $\gamma$, and energy fluxes, $F_E$, along the jets: while the jet in model S-T60 accelerates to $\gamma\sim7$ and conserves its energy flux, the jet in model W-T60 loses energy and fails to accelerate to relativistic speeds at early times when the phase lag is large.

\begin{figure}
  \begin{center}
    \includegraphics[width=\linewidth,trim=0mm
    0mm 0mm 0,clip]{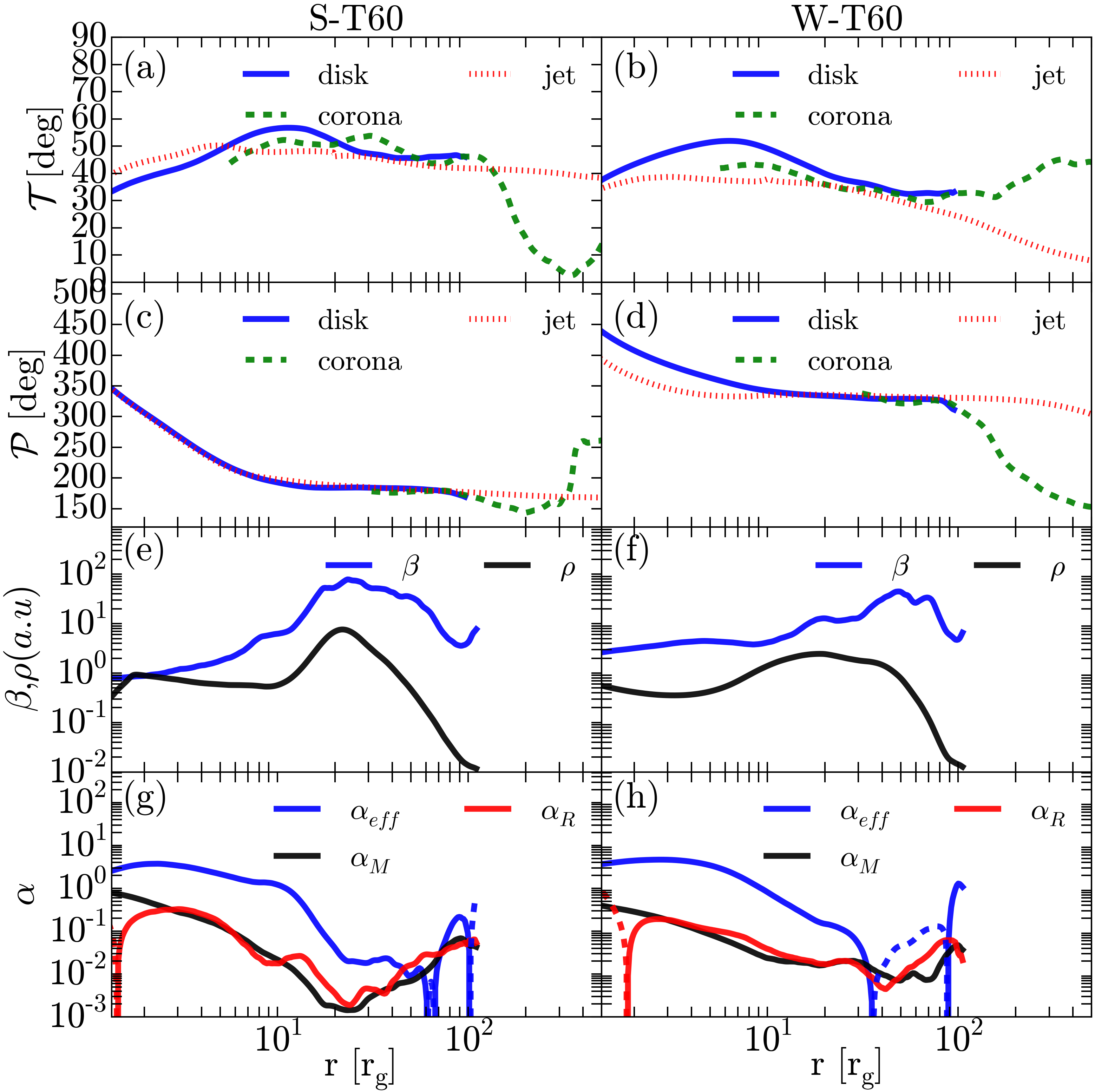}
  \end{center}
  \caption{The tilt angle $\mathcal{T}$ in panels (a,b) and precession $\mathcal{P}$ angle in panels (c,d) as function of radius. Strong flux model S-T60 (left panel) is averaged between $3.5\times10^4<t<4.0\times10^4$ $r_g/c$ and weak flux model W-T60 (right panel) is averaged between $5.5\times10^4<t<6.0\times10^4$ $r_g/c$. [panels (a,b)]: The disc does not align with the BH \citep{bp75}, and instead develops what looks like a radial tilt oscillation in $\mathcal{T}$ peaking at $r \approx 10r_g$ (see the main text). [panels (c,d)]: Disc precession angle $\mathcal{P}$ decreases as (positive) function of radius for $r\lesssim 10r_g$ reflecting the warping of the inner disc. In panel (d), outside the disc edge at $r\simeq 100r_g$, coronal precession angle sharply drops off as (positive) function of radius, reflecting the precessional lag of the corona behind the disc. [panels (e,f)]: Density, $\rho$, and gas-to-magnetic pressure ratio, $\beta$, are similar for both models. In (g,h) the Maxwell, $\alpha_M$, and Reynolds, $\alpha_R$, stresses are much higher within the inner disc ($r<10$ $r_g$) with respect to the outer disk ($r>10$ $r_g$), while their sum is insufficient to account for the effective viscosity $\alpha_{\rm eff}$ (Sec.~\ref{sec:Results}).}
\label{fig:radplot}
\end{figure}

Figure~\ref{fig:tiltvst}(a,b), shows the time evolution of net tilt $\mathcal T$ and precession $\mathcal P$ angles (see \citep{Liska2018A, Liska2018C} for definitions) for the disc, corona and jets, allowing us to quantitatively study the phase lags. 
We caution that the precession frequency depends sensitively on the size of the disc, which is ultimately set by the simulation setup. However, since disc precession period can be estimated by integrating the total LT torque (e.g. \citealt{Fragile2007,Liska2018A}), a thick disc of such a size is expected in some models for the hard state \citep{Esin1997}. In both of our models, $\mathcal P$ increases quasi-monotonically at approximately the same rate, reaching $200^{\circ}{-}400^{\circ}$. This differs from our previous work \citep{Liska2018A}, where the disc stopped precessing at $\mathcal P \simeq 40^{\circ}$ due to very fast viscous expansion of the thicker, $h/r=0.3$, disc (see also Sec.~\ref{sec:introduction}). The phase lag between the disc and corona, $\Delta\mathcal{P}$, in Fig.~\ref{fig:tiltvst}(e,f) initially peaks at $\Delta\mathcal{P}\sim50^{\circ}$ before stabilizing at $\Delta\mathcal{P}\sim10^{\circ}$.  There is a strong correlation between the phase lag and jet power (Fig.~\ref{fig:time}(e,f)), suggesting that the corona does not only disrupt the jet (if it is weak) when the phase lag is large, but that the initially disc-aligned jet can also torque the corona into (partial) alignment with itself. In both models the disc-corona-jet system  slowly evolves as a whole  toward alignment with the BH on the viscous timescale, from the initial $\mathcal T = 60^\circ$ to $\mathcal T\sim30^\circ$ at $t=10^5r_g/c$. This global alignment mode differs (see \citep{Liska2018C}), from the Bardeen\&Petterson alignment \citep{bp75}, which only affects the inner disc and occurs on time-scales much shorter than the viscous timescale of the outer disc. Both alignment modes might be caused due to the turbulent mixing (and cancellation) of misaligned angular momentum in a warp \citep{Sorathia2013}, especially when the precession angle shows a steep dependence on radius (Fig.~\ref{fig:radplot}(c,d)). Since we have a finitely sized disc and this global alignment happens on the viscous time we expect that this global alignment mode will disappear if the disc is fed by misaligned angular momentum from an outer disc.

Figure~\ref{fig:radplot}(a--d) shows the radial runs of tilt and precession angles of the disc, corona, and jet. Disc tilt increases away from the BH, peaks at $\sim 10r_g$ and drops thereafter. While this qualitatively agrees with the analytic theory for the bending wave regime
large-tilt non-linear effects might introduce corrections \citep{White2019A}. Furthermore, the disc, corona and jet are (roughly) aligned with each other up to the disc's outer edge located at $\sim100r_g$, which is consistent with previous work \citep{Liska2018A}. At $r \sim 100 r_g$, corona's precession angle sharply drops, which is qualitatively consistent with the phase lag discussed above. At $r\gtrsim200 r_g$, corona's tilt and precession angles show large amplitude oscillations, likely because the corona does not rotate in a single plane anymore and is partially outflowing, leading to these angles being ill-defined. However, since most of the corona's angular momentum lies at $r\lesssim200 r_g$, this region does not significantly contribute to the net tilt and precession angles shown in Fig.~\ref{fig:tiltvst}(a--d).

Figure ~\ref{fig:radplot}(e,f) shows that both models remain gas pressure dominated,$\beta=\langle p_g\rangle_\rho/\langle p_m\rangle_\rho>1$ down to the event horizon, where $\langle q\rangle_\rho = \int \rho q\,{\rm d}A/\int \rho\,{\rm d}A$ is density-weighted average and ${\rm d}A$ is the surface element. The Maxwell ($\alpha_M=b^r b^{\phi}/(p_g+p_b)$, $b^\mu$ is the magnetic 4-vector) and Reynolds ($\alpha_R=\rho u^r u^{\phi}/(p_g+p_b)$, $u^\mu$ is the 4-velocity) stresses in Fig.~\ref{fig:radplot}(g,h), calculated in a coordinate system aligned with the local angular momentum, are similar. However, their sum does not match the effective viscosity ($\alpha_{\rm eff}=-v_r v_k/ c_{s}^2$, where $v_k$ is the Keplerian 3-velocity and $c_s$ is the sound speed). Pending future analysis this may be indicative of pressure gradients, large scale torques driving outflows, which remove the disc's angular momentum (similar to \citealt{Liska2018C}), warps and/or spiral shocks that form in tilted discs (e.g. \citealt{Nelson2000, Fragile2008, Liska2022B, Kaaz2022B}).

\section{Discussion and Conclusion}
\label{sec:Conclusion}
In this work we present GRMHD simulations of highly tilted ($60$ degrees), moderately thin ($h/r \approx 0.1$) accretion discs around rapidly spinning BHs ($a=0.9375$). We show that such discs produce precessing relativistic jets. The disc-jet system completes the full precession cycle under the action of Lense-Thirring torques. This does not happen in thicker discs ($h/r\sim0.3$) where viscous spreading causes the precession to stall before the disc is able complete a full precession period \citep{Liska2018A}. The precession-induced quasi-periodic variability in XRB lightcurves could potentially be exploited to measure BH spin \citep{Stella1998, Ingram2009, Musoke2022}. However, this cannot be done for the discs presented in this article since their precession period is influenced by the size of the disc in the initial conditions. One way a precessing disc can form more self-consistently is by tearing up a large misaligned accretion disc: the disc tears up at a radius where the Lense-Thirring torques, which try to tear the disc apart, overwhelm the viscous torques, which hold the disc together \citep{Nixon2012B,Nealon2015, Liska2021}. Following disc tearing the size of the disc and hence the precession period (for a given BH spin) are determined self-consistently (among others) by the thickness of the disc, the misalignment angle, and the amount of magnetic flux in the disc. However, disc tearing was only observed in GRMHD simulations with scaleheight $h/r \lesssim 0.03 $ \citep{Liska2021}, which is significantly smaller than the $h/r=0.1$ discs presented in this article.

\begin{figure}
\begin{center}
\includegraphics[width=\linewidth,trim=0mm 0mm 0mm 0,clip]{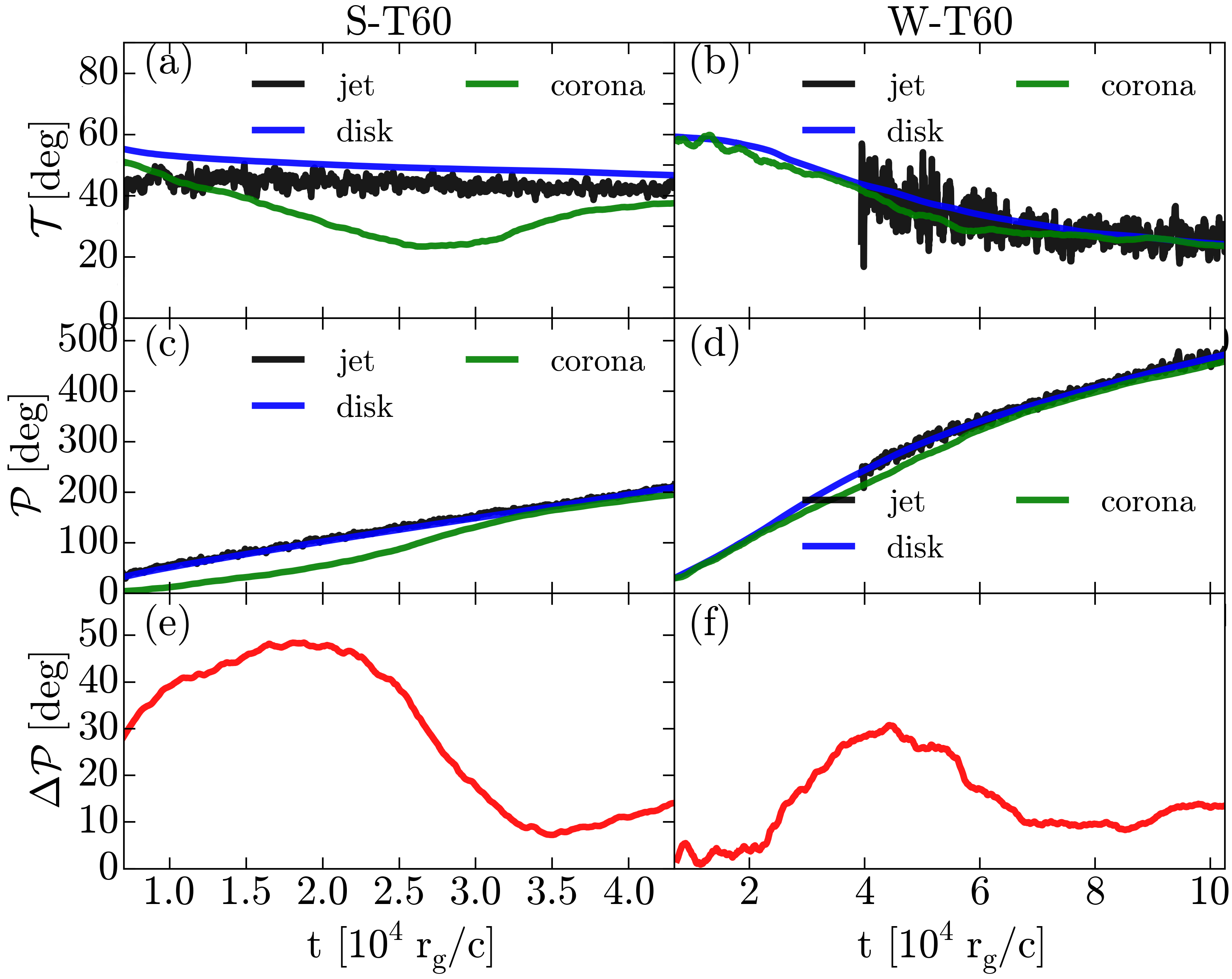}
\end{center}
\caption{The tilt angle, $\mathcal{T}$, in (a,b) and precession angle, $\mathcal{P}$, in (c,d) as functions of time, for the disc, corona, and jet. The disc's and corona's tilt and precession angles are obtained from their total angular momentum \citep{Liska2018A}. The jet's tilt and precession angle is obtained from the energy density weighted centroid position ($x_{jet}, y_{jet}, z_{jet}$) \citep{Liska2018A}. Model S-T60 and W-T60 both align with the BH over time, while their precession angle grows semi-monotonically. The phase lag, $\Delta \mathcal{P}$, between disc and corona in (e,f) first peaks, but seems to stabilize later at a lower value. This suggests that the disc and corona experience periods of differential precession in both directions with a net increase their relative precession angle.}
\label{fig:tiltvst}
\end{figure}

We found that the magnetic pressure dominated disc wind, which we refer to as the corona, lags the disc in phase by $10{-}40^\circ$ (Fig.~\ref{fig:tiltvst}c). Recent two-temperature radiative GRMHD simulations \citep{Liska2022} suggest that such disc winds launched from thin accretion discs can reach electron temperatures of $T_e \gtrsim 5 \times 10^8K$. This can help to interpret the phase lags of Type-C QPOs observed in XRBs \citep{Wijnands1999, Reig2000, Qu2010, Pahari2013, Eijnden2016A, Eijnden2017}. More specifically, \citet{Eijnden2016A} recently used a time-resolved cross-spectral analysis of GRS 1915+105 and found that the phase lags systematically change on short timescale in a way indicating that  for observations with $\nu_{\rm QPO}>2$ Hz for $5$-$10$ cycles the QPO in the hard band is slightly faster than that in the soft band, resulting in a gradually increasing soft lag, and for observations with $\nu_{\rm QPO}<2$ Hz the QPO in the hard band is slightly slower than that in the soft band, resulting in a gradually increasing hard lag. The increase in soft lag for $\nu_{\rm QPO}>2$ Hz was suggested to result from differential precession, since inner regions are expected to have a harder spectrum and to precess faster than outer regions \citep{Stella1998}. However, the increase in hard lag for $\nu_{\rm QPO}<2$ Hz over a QPO period remained a mystery.
The evolving coronal lag in our simulations (i.e. the increase in $\Delta\mathcal{P}$) may provide a natural solution to this puzzle: the corona, which emits a harder spectrum than the disc (via inverse Compton scattering), precesses more slowly than the disc, resulting in a faster QPO in the soft band. Translating the coronal lag into X-ray lightcurves is beyond the scope of this work. However, the timescale for the phase lag ($\lesssim 1$ precession period) build up and its magnitude ($10{-}40^\circ$) match observations \citep{Eijnden2016A} reasonably well. Our simulations also provide some clues why for $\nu_{\rm QPO}>2$ Hz the phase lag turns soft. As $\nu_{\rm QPO}$ increases, one would expect the precessing inner disc to become smaller \citep{Stella1998,Ingram2009}. If the hardness of the corona with respect to the disc decreases as the disc becomes smaller the phase lag will gradually shift from hard to soft. However, this will need to be tested in future work that properly models the heating/cooling and emission of both the disc and corona.

 While the strong jet in S-T60 propagates unhindered, the weak jet in W-T60 disrupts and stalls at $r\lesssim 10^2 r_g$, obstructed by the lagging corona (Figs.~\ref{fig:contourplot}, \ref{fig:jetplot}). We expect that a precessing jet is more likely to become truncated by a given ambient medium than a non-precessing jet because it will need to clear out a larger volume of surrounding material. Such truncated jets may transfer significant amounts of energy and angular momentum to the ambient medium and, by doing so, contribute to weakly collimated matter-dominated sub-relativistic outflows. In addition, the jet-ambient medium interaction might lead to magnetic reconnection and acceleration of (non-)thermal particles, possibly producing flares across the electromagnetic spectrum. However, if such jets are truncated at small radii, they might be difficult to detect. Future work will need to determine the observational signatures of such truncated jets and address all of the physics (in addition to precession) responsible for jet truncation.

We have shown that the jets can be deflected by the ambient medium, which could be produced by the same accretion system that produced the jet. Whether such a deflection is possible depends on the jet power. If the jet is powerful enough, the jet will plow through and will be hardly deflected. On the other hand, if the jet is weak, it will be torqued into alignment with the surrounding medium and, as in model W-T60, may stall. For instance, even though relativistic jets derive their power from the BH spin, they do not have to point along the BH rotational axis. In particular, changes in jet orientation observed in AGN do not imply the changes in the BH spin direction.

\section{Acknowledgments}
We thank Marta Volonteri for useful feedback. This research was made possible by NSF PRAC award no. 1615281 and OAC-1811605 at the
Blue Waters sustained-petascale computing project. ML and MK were supported by the NWO Spinoza Prize, AI by the Royal Society URF, and SM by the NWO VICI grant (no. 639.043.513).

\section{Supporting Information}
Additional Supporting Information may be found in the online version of this article: movie files. See our \href{https://www.youtube.com/playlist?list=PLDO1oeU33GwmwOV_Hp9s7572JdU8JPSSK}{YouTube playlist} for 3D visualizations of all models.

\label{sec:acks}

{\small
\bibliography{mybib,sasha,newbib}

\begin{thebibliography}{}

\bibitem[\protect\citeauthoryear{{Balbus} \& {Hawley}}{{Balbus} \&
  {Hawley}}{1991}]{Balbus1991}
{Balbus} S.~A.,  {Hawley} J.~F.,  1991, \apj, 376, 214

\bibitem[\protect\citeauthoryear{Balbus \& Hawley}{Balbus \&
  Hawley}{1998}]{Balbus1998}
Balbus S.~A.,  Hawley J.~F.,  1998, Rev. Mod. Phys., 70, 1

\bibitem[\protect\citeauthoryear{{Bardeen} \& {Petterson}}{{Bardeen} \&
  {Petterson}}{1975}]{bp75}
{Bardeen} J.~M.,  {Petterson} J.~A.,  1975, \apjl, 195, L65

\bibitem[\protect\citeauthoryear{{Beloborodov}}{{Beloborodov}}{1999}]{Beloborodov1999}
{Beloborodov} A.~M.,  1999, \apjl, 510, L123

\bibitem[\protect\citeauthoryear{{Caproni}, {Abraham}, {Livio} \& {Mosquera
  Cuesta}}{{Caproni} et~al.}{2007}]{Caproni2007}
{Caproni} A.,  {Abraham} Z.,  {Livio} M.,    {Mosquera Cuesta} H.~J.,  2007,
  \mnras, 379, 135

\bibitem[\protect\citeauthoryear{{Caproni}, {Abraham} \& {Mosquera
  Cuesta}}{{Caproni} et~al.}{2006}]{Caproni2006}
{Caproni} A.,  {Abraham} Z.,    {Mosquera Cuesta} H.~J.,  2006, \apj, 638, 120

\bibitem[\protect\citeauthoryear{{Esin}, {McClintock} \& {Narayan}}{{Esin}
  et~al.}{1997}]{Esin1997}
{Esin} A.~A.,  {McClintock} J.~E.,    {Narayan} R.,  1997, \apj, 489, 865

\bibitem[\protect\citeauthoryear{{Ferreira}, {Petrucci}, {Henri}, {Saug{\'e}}
  \& {Pelletier}}{{Ferreira} et~al.}{2006}]{Fereira2006}
{Ferreira} J.,  {Petrucci} P.~O.,  {Henri} G.,  {Saug{\'e}} L.,    {Pelletier}
  G.,  2006, \aap, 447, 813

\bibitem[\protect\citeauthoryear{{Fishbone} \& {Moncrief}}{{Fishbone} \&
  {Moncrief}}{1976}]{Fishbone1976}
{Fishbone} L.~G.,  {Moncrief} V.,  1976, \apj, 207, 962

\bibitem[\protect\citeauthoryear{{Fragile} \& {Blaes}}{{Fragile} \&
  {Blaes}}{2008}]{Fragile2008}
{Fragile} P.~C.,  {Blaes} O.~M.,  2008, \apj, 687, 757

\bibitem[\protect\citeauthoryear{{Fragile}, {Blaes}, {Anninos} \&
  {Salmonson}}{{Fragile} et~al.}{2007}]{Fragile2007}
{Fragile} P.~C.,  {Blaes} O.~M.,  {Anninos} P.,    {Salmonson} J.~D.,  2007,
  \apj, 668, 417

\bibitem[\protect\citeauthoryear{{Greene}, {Bailyn} \& {Orosz}}{{Greene}
  et~al.}{2001}]{Greene2001}
{Greene} J.,  {Bailyn} C.~D.,    {Orosz} J.~A.,  2001, \apj, 554, 1290

\bibitem[\protect\citeauthoryear{{Greene}, {Seth}, {den Brok}, {Braatz},
  {Henkel}, {Sun}, {Peng}, {Kuo}, {Impellizzeri} \& {Lo}}{{Greene}
  et~al.}{2013}]{Greene2013}
{Greene} J.~E.,  {Seth} A.,  {den Brok} M.,  {Braatz} J.~A.,  {Henkel} C.,
  {Sun} A.-L.,  {Peng} C.~Y.,  {Kuo} C.-Y.,  {Impellizzeri} C.~M.~V.,    {Lo}
  K.~Y.,  2013, \apj, 771, 121

\bibitem[\protect\citeauthoryear{{Hawley}, {Richers}, {Guan} \&
  {Krolik}}{{Hawley} et~al.}{2013}]{2013ApJ...772..102H}
{Hawley} J.~F.,  {Richers} S.~A.,  {Guan} X.,    {Krolik} J.~H.,  2013, \apj,
  772, 102

\bibitem[\protect\citeauthoryear{{Herrnstein}, {Moran}, {Greenhill} \&
  {Trotter}}{{Herrnstein} et~al.}{2005}]{Herrnstein2005}
{Herrnstein} J.~R.,  {Moran} J.~M.,  {Greenhill} L.~J.,    {Trotter} A.~S.,
  2005, \apj, 629, 719

\bibitem[\protect\citeauthoryear{{Hjellming} \& {Rupen}}{{Hjellming} \&
  {Rupen}}{1995}]{hjellming95}
{Hjellming} R.~M.,  {Rupen} M.~P.,  1995, \nat, 375, 464

\bibitem[\protect\citeauthoryear{{Ingram}, {Done} \& {Fragile}}{{Ingram}
  et~al.}{2009}]{Ingram2009}
{Ingram} A.,  {Done} C.,    {Fragile} P.~C.,  2009, \mnras, 397, L101

\bibitem[\protect\citeauthoryear{{Ivanov} \& {Illarionov}}{{Ivanov} \&
  {Illarionov}}{1997}]{Ivanov1997}
{Ivanov} P.~B.,  {Illarionov} A.~F.,  1997, \mnras, 285, 394

\bibitem[\protect\citeauthoryear{{Kaaz}, {Liska}, {Jacquemin-Ide}, {Andalman},
  {Musoke}, {Tchekhovskoy} \& {Porth}}{{Kaaz} et~al.}{2022}]{Kaaz2022B}
{Kaaz} N.,  {Liska} M. T.~P.,  {Jacquemin-Ide} J.,  {Andalman} Z.~L.,  {Musoke}
  G.,  {Tchekhovskoy} A.,    {Porth} O.,  2022, arXiv e-prints, p.
  arXiv:2210.10053

\bibitem[\protect\citeauthoryear{{King}, {Lubow}, {Ogilvie} \&
  {Pringle}}{{King} et~al.}{2005}]{King2005}
{King} A.~R.,  {Lubow} S.~H.,  {Ogilvie} G.~I.,    {Pringle} J.~E.,  2005,
  \mnras, 363, 49

\bibitem[\protect\citeauthoryear{{Lense} \& {Thirring}}{{Lense} \&
  {Thirring}}{1918}]{lense18}
{Lense} J.,  {Thirring} H.,  1918, Physikalische Zeitschrift, 19

\bibitem[\protect\citeauthoryear{{Liska}, {Chatterjee}, {Tchekhovskoy}, {Yoon},
  {van Eijnatten}, {Hesp}, {Markoff}, {Ingram} \& {van der Klis}}{{Liska}
  et~al.}{2019}]{Liska2020}
{Liska} M.,  {Chatterjee} K.,  {Tchekhovskoy} A.,  {Yoon} D.,  {van Eijnatten}
  D.,  {Hesp} C.,  {Markoff} S.,  {Ingram} A.,    {van der Klis} M.,  2019,
  arXiv e-prints, p. arXiv:1912.10192

\bibitem[\protect\citeauthoryear{{Liska}, {Hesp}, {Tchekhovskoy}, {Ingram},
  {van der Klis} \& {Markoff}}{{Liska} et~al.}{2018}]{Liska2018A}
{Liska} M.,  {Hesp} C.,  {Tchekhovskoy} A.,  {Ingram} A.,  {van der Klis} M.,
   {Markoff} S.,  2018, \mnras, 474, L81

\bibitem[\protect\citeauthoryear{{Liska}, {Hesp}, {Tchekhovskoy}, {Ingram},
  {van der Klis}, {Markoff} \& {Van Moer}}{{Liska} et~al.}{2021}]{Liska2021}
{Liska} M.,  {Hesp} C.,  {Tchekhovskoy} A.,  {Ingram} A.,  {van der Klis} M.,
  {Markoff} S.~B.,    {Van Moer} M.,  2021, \mnras, 507, 983

\bibitem[\protect\citeauthoryear{{Liska}, {Tchekhovskoy}, {Ingram} \& {van der
  Klis}}{{Liska} et~al.}{2019}]{Liska2018C}
{Liska} M.,  {Tchekhovskoy} A.,  {Ingram} A.,    {van der Klis} M.,  2019,
  \mnras, p.~813

\bibitem[\protect\citeauthoryear{{Liska}, {Kaaz}, {Musoke}, {Tchekhovskoy} \&
  {Porth}}{{Liska} et~al.}{2022}]{Liska2022B}
{Liska} M.~T.~P.,  {Kaaz} N.,  {Musoke} G.,  {Tchekhovskoy} A.,    {Porth} O.,
  2022, arXiv e-prints, p. arXiv:2210.10198

\bibitem[\protect\citeauthoryear{{Liska}, {Musoke}, {Tchekhovskoy}, {Porth} \&
  {Beloborodov}}{{Liska} et~al.}{2022}]{Liska2022}
{Liska} M.~T.~P.,  {Musoke} G.,  {Tchekhovskoy} A.,  {Porth} O.,
  {Beloborodov} A.~M.,  2022, arXiv e-prints, p. arXiv:2201.03526

\bibitem[\protect\citeauthoryear{{Lubow}, {Ogilvie} \& {Pringle}}{{Lubow}
  et~al.}{2002}]{Lubow2002}
{Lubow} S.~H.,  {Ogilvie} G.~I.,    {Pringle} J.~E.,  2002, \mnras, 337, 706

\bibitem[\protect\citeauthoryear{{Maccarone}}{{Maccarone}}{2002}]{Maccarone2002}
{Maccarone} T.~J.,  2002, \mnras, 336, 1371

\bibitem[\protect\citeauthoryear{{Musoke}, {Liska}, {Porth}, {van der Klis} \&
  {Ingram}}{{Musoke} et~al.}{2022}]{Musoke2022}
{Musoke} G.,  {Liska} M.,  {Porth} O.,  {van der Klis} M.,    {Ingram} A.,
  2022, arXiv e-prints, p. arXiv:2201.03085

\bibitem[\protect\citeauthoryear{{Narayan} \& {Yi}}{{Narayan} \&
  {Yi}}{1994}]{Narayan1994}
{Narayan} R.,  {Yi} I.,  1994, \apjl, 428, L13

\bibitem[\protect\citeauthoryear{{Nealon}, {Price} \& {Nixon}}{{Nealon}
  et~al.}{2015}]{Nealon2015}
{Nealon} R.,  {Price} D.~J.,    {Nixon} C.~J.,  2015, \mnras, 448, 1526

\bibitem[\protect\citeauthoryear{{Nelson} \& {Papaloizou}}{{Nelson} \&
  {Papaloizou}}{2000}]{Nelson2000}
{Nelson} R.~P.,  {Papaloizou} J.~C.~B.,  2000, \mnras, 315, 570

\bibitem[\protect\citeauthoryear{{Nixon}, {King}, {Price} \& {Frank}}{{Nixon}
  et~al.}{2012}]{Nixon2012B}
{Nixon} C.,  {King} A.,  {Price} D.,    {Frank} J.,  2012, \apjl, 757, L24

\bibitem[\protect\citeauthoryear{{Noble}, {Krolik} \& {Hawley}}{{Noble}
  et~al.}{2009}]{Noble2009}
{Noble} S.~C.,  {Krolik} J.~H.,    {Hawley} J.~F.,  2009, \apj, 692, 411

\bibitem[\protect\citeauthoryear{{Novikov} \& {Thorne}}{{Novikov} \&
  {Thorne}}{1973}]{Novikov1973}
{Novikov} I.~D.,  {Thorne} K.~S.,  1973, in {Dewitt} C.,  {Dewitt} B.~S.,  eds,
  Black Holes (Les Astres Occlus) {Astrophysics of black holes.}.
pp 343--450

\bibitem[\protect\citeauthoryear{Pahari, Neilsen, Yadav, Misra \&
  Uttley}{Pahari et~al.}{2013}]{Pahari2013}
Pahari M.,  Neilsen J.,  Yadav J.~S.,  Misra R.,    Uttley P.,  2013, ApJ, 778,
  136

\bibitem[\protect\citeauthoryear{{Papaloizou} \& {Lin}}{{Papaloizou} \&
  {Lin}}{1995}]{papaloizou95}
{Papaloizou} J.~C.~B.,  {Lin} D.~N.~C.,  1995, \araa, 33, 505

\bibitem[\protect\citeauthoryear{{Papaloizou} \& {Pringle}}{{Papaloizou} \&
  {Pringle}}{1983}]{Papaloizou1983}
{Papaloizou} J.~C.~B.,  {Pringle} J.~E.,  1983, \mnras, 202, 1181

\bibitem[\protect\citeauthoryear{Qu, Lu, Lu, Song, Zhang, Ding \& Wang}{Qu
  et~al.}{2010}]{Qu2010}
Qu J.~L.,  Lu F.~J.,  Lu Y.,  Song L.~M.,  Zhang S.,  Ding G.~Q.,    Wang
  J.~M.,  2010, ApJ, 710, 836

\bibitem[\protect\citeauthoryear{Reig, Belloni, van~der Klis, Méndez, Kylafis
  \& Ford}{Reig et~al.}{2000}]{Reig2000}
Reig P.,  Belloni T.,  van~der Klis M.,  Méndez M.,  Kylafis N.~D.,    Ford
  E.~C.,  2000, ApJ, 541, 883

\bibitem[\protect\citeauthoryear{{Ressler}, {Tchekhovskoy}, {Quataert},
  {Chandra} \& {Gammie}}{{Ressler} et~al.}{2015}]{Ressler2015}
{Ressler} S.~M.,  {Tchekhovskoy} A.,  {Quataert} E.,  {Chandra} M.,    {Gammie}
  C.~F.,  2015, \mnras, 454, 1848

\bibitem[\protect\citeauthoryear{{Sorathia}, {Krolik} \& {Hawley}}{{Sorathia}
  et~al.}{2013}]{Sorathia2013}
{Sorathia} K.~A.,  {Krolik} J.~H.,    {Hawley} J.~F.,  2013, \apj, 777, 21

\bibitem[\protect\citeauthoryear{{Stella} \& {Vietri}}{{Stella} \&
  {Vietri}}{1998}]{Stella1998}
{Stella} L.,  {Vietri} M.,  1998, \apjl, 492, L59

\bibitem[\protect\citeauthoryear{{van den Eijnden}, {Ingram} \& {Uttley}}{{van
  den Eijnden} et~al.}{2016}]{Eijnden2016A}
{van den Eijnden} J.,  {Ingram} A.,    {Uttley} P.,  2016, \mnras, 458, 3655

\bibitem[\protect\citeauthoryear{{van den Eijnden}, {Ingram}, {Uttley},
  {Motta}, {Belloni} \& {Gardenier}}{{van den Eijnden}
  et~al.}{2017}]{Eijnden2017}
{van den Eijnden} J.,  {Ingram} A.,  {Uttley} P.,  {Motta} S.~E.,  {Belloni}
  T.~M.,    {Gardenier} D.~W.,  2017, \mnras, 464, 2643

\bibitem[\protect\citeauthoryear{van~der Klis}{van~der Klis}{1989}]{Klis1989}
van~der Klis M.,  1989, Annual Review of Astronomy and Astrophysics, 27, 517

\bibitem[\protect\citeauthoryear{{Volonteri}, {Madau}, {Quataert} \&
  {Rees}}{{Volonteri} et~al.}{2005}]{Volonteri2005}
{Volonteri} M.,  {Madau} P.,  {Quataert} E.,    {Rees} M.~J.,  2005, \apj, 620,
  69

\bibitem[\protect\citeauthoryear{{White}, {Quataert} \& {Blaes}}{{White}
  et~al.}{2019}]{White2019A}
{White} C.~J.,  {Quataert} E.,    {Blaes} O.,  2019, arXiv e-prints, p.
  arXiv:1902.09662

\bibitem[\protect\citeauthoryear{{Wijnands}, {Homan} \& {van der
  Klis}}{{Wijnands} et~al.}{1999}]{Wijnands1999}
{Wijnands} R.,  {Homan} J.,    {van der Klis} M.,  1999, \apjl, 526, L33

\end{thebibliography}
\bibliographystyle{mn2e}

}
\label{lastpage}
\end{document}